\documentstyle[aps,preprint,eqsecnum]{revtex}
\begin{document}
\draft
\preprint{$\hbox to 5 truecm{\hfill Alberta-Thy-39-94}\atop
{\hbox to 5 truecm{\hfill gr-qc/9412051}\atop
\hbox to 5 truecm{\hfill December 1994}}$}
\title{Microcanonical functional integral and entropy for \\
eternal black holes}
\author{Erik A. Martinez
\footnote{electronic address: martinez@phys.ualberta.ca}}
\address{Theoretical Physics Institute,\\
University of Alberta,\\
Edmonton, Alberta T6G 2J1, Canada.}
\maketitle
\begin{abstract}
\noindent{\baselineskip=16pt
The microcanonical functional integral for an eternal black hole system is
considered.  This requires computing the microcanonical action
for a spatially bounded spacetime region when its two
disconnected  timelike boundary surfaces are located in different wedges of
the Kruskal diagram. The path integral is a sum over Lorentzian geometries
and is evaluated semiclassically when its boundary data are chosen such
that the
system is approximated by any Lorentzian, stationary  eternal black hole.
This approach opens the possibility of including explicitly the internal
degrees of  freedom of a physical black hole in path integral descriptions of
its thermodynamical properties. If the functional integral is interpreted as
the density of states of the system, the corresponding entropy equals
${\cal S} = A_H/4 - A_H/4 = 0$ in the semiclassical approximation, where $A_H$
is the area of the black hole horizon. The functional integral reflects the
properties of a pure state.  The description of the black hole density of
states
in terms of the eternal black hole functional integral is also discussed.
\par}
\end{abstract}
\vspace{7mm}
\vfill
\eject

\section{Introduction}

Despite considerable progress in the path integral description of
gravitational systems \cite{BrYo2,BrYoRev,Te,CaTe,HaHoRo,BrMaYo}, the
statistical mechanical origin of  black hole entropy remains unclear in this
approach. The dynamical origin of entropy has been recently studied with the
help of different methods (see, for example, Refs.\cite{FrNo,BaFrZe,LoWh}).
Given these developments, it would be interesting to  include explicitly the
internal degrees of freedom of a  black hole  in the functional integral
and study their contribution to black hole entropy. An attempt in this
direction, which we pursue in this paper, consists in investigating the
microcanonical functional integral when applied to an eternal black hole
statistical system which effectively contains information about the internal
degrees of freedom of a physical black hole.

A proposal for the density of states of a gravitational
system obtained as the trace of a microcanonical density matrix has been
suggested  recently in Refs. \cite{BrYo2,BrYoRev}.  The density of quantum
states for a  self-gravitating system  spatially bounded by a timelike
three-dimensional surface $B$  is given by the functional
integral
\begin{equation}
\nu[\varepsilon,j,\sigma] =
\sum_{\sl M} \int {\cal D}H \exp(i S_m/\hbar) \ . \label{nu}
\end{equation}
The phase of the functional integral is proportional to the so-called
microcanonical action $S_m$ which describes the dynamics of a gravitational
system whose  surface energy density
$\varepsilon$, surface momentum density $j_a$, and size (specified by the
two-dimensional metric $\sigma_{ab}$) are fixed at the spatial boundary.
 The quantities $\varepsilon$, $j_a$, and $\sigma_{ab}$ are constructed from
the dynamical phase space variables that include the three-metric $h_{ij}$ of
an
initial spacelike  hypersurface $\Sigma$ and its conjugate momentum $P^{ij}$.
The density of states  is  defined as a formal integral over
Lorentzian metrics that satisfy the boundary conditions and is a
functional of the quantities
 $\varepsilon$, $j_a$, and $\sigma_{ab}$.
The sum over $M$
in (\ref{nu}) refers to a sum over manifolds of different topologies
which are periodic in the time-like direction and whose three-dimensional
boundary has topology $S \times S^1$, where the two-dimensional surface $S$
is the intersection of the boundary  $B$ and the slice
$\Sigma$.   The symbol ${\cal D}H$ in (\ref{nu}) denotes a formal measure in
the space of these manifolds.
The black hole density of states $\nu_{*}$ is obtained from the functional
integral (\ref{nu}) when the latter is approximated semiclassically by using a
complex metric whose boundary data at its single boundary surface coincide with
the boundary data of a Lorentzian, stationary, axisymmetric black hole.
 The density of states defined accordingly equals the exponential of
one fourth of the area of the black hole horizon.

The proposal (\ref{nu}) opens the possibility of determining the
thermodynamical properties of black hole systems starting from a sum
over real Lorentzian geometries.
However, several problems remain in this approach.  First,  a spacelike
hypersurface $\Sigma$ that describes the initial data of a  Lorentzian black
hole  has to cross necessarily the event horizon and eventually intersect the
interior singularity.
This implies that additional information has to be provided on $\Sigma$ in
order to describe the properties of the singularity \cite{FrMa2}.
Second,  the microcanonical functional integral and
action used in \cite{BrYo2} to calculate the black hole entropy are
appropriate when the spacetime has a single  timelike boundary surface.
However, as already noted in  \cite{BrYo2},  a  Lorentzian, stationary,
axisymmetric black hole is not a extremum of this action since  it cannot be
placed on a manifold with a single timelike boundary.
In particular, this implies that the black hole density of states $\nu_{*}$
whose boundary data correspond to the boundary data  of a   Lorentzian,
stationary, axisymmetric black hole cannot be approximated semiclassically  by
using the same Lorentzian metric that motivates its boundary conditions.

These difficulties do not prevent the evaluation of the black hole
density of states in the semiclassical approximation \cite{BrYo2}. As already
mentioned, there exists a related complex metric which satisfies the boundary
data and which can be used to calculate the Lorentzian functional integral in a
steepest descents approximation by distorting its contours of
integration \cite{BrMaYo,BrYo2}. This approximation yields the correct result
for the black hole entropy but conceals its origin.  As with other
complexification schemes  previously used in calculations of black hole
partition functions \cite{GiHa1,Yo,WhYo},  the interior of the Lorentzian black
hole literally disappears by virtue of
this procedure, leaving effectively only a
periodically identified Euclidean version of the  ``right" wedge region of a
Kruskal diagram. The properties of the black hole interior become encoded  in a
set of conditions at the so-called ``bolt" of the complex geometry
\cite{GiHa2}.
In this approach, as in other formulations of gravitational thermodynamics in
terms of  path integrals, the statistical origin of entropy and its
relationship
to the internal degrees of freedom of a black hole remain obscure.

We believe that the  problems mentioned above and the role of internal
degrees of freedom in  functional integral descriptions of black hole
thermodynamics can be addressed by explicitly
considering the  eternal version of a black hole.
The description of states of a physical black hole
formed from gravitational collapse in terms of the  states of
its eternal version  has been proposed in Ref.\cite{BaFrZe}. The late time
geometry of a physical black hole can be
analytically continued into the spacetime
of an eternal black hole if the latter configuration possesses the same
macroscopic parameters as the former one. The excitations of the physical black
hole can be associated with the deformations of an initial global Cauchy
surface $\Sigma$ of the eternal black hole plus initial data for the
non-gravitational fields defined on such a distorted surface
\cite{BaFrZe,FrMa2}. In general, the spatial slices $\Sigma$ that foliate an
eternal black hole are (deformed) Einstein-Rosen bridges with wormhole topology
$R^1 \times S^2$. The spacetime is composed of two  wedges  $M_+$ and $M_-$
located in  the right ($R_+$) and left ($R_-$) sectors of a Kruskal diagram
\cite{FrMa2}. Internal and external degrees of freedom of the black hole can be
easily identified in this approach since the hypersurfaces $\Sigma$ are
naturally divided in two parts $\Sigma_+$ and $\Sigma_-$ by a bifurcation
two-surface $S_0$.   While the  ``external" degrees of freedom of the original
black hole are naturally given by the initial data at $\Sigma_+$,  its
``internal"  degrees of freedom can be identified with initial data defined at
$\Sigma_-$.

The importance of finite size systems in gravitational thermodynamics has
been stressed repeatedly \cite{Yo,BrYoRev}.
Finite spacetime regions are required in thermodynamical applications since a
gravitational system in thermal equilibrium with a radiation bath is not
described by an asymptotically flat spacetime. In particular, rotating black
holes can  be in thermal equilibrium only if contained inside a spatially
finite
boundary \cite{IsSt,FrTh,BrMaYo}.
Other advantages of bounded systems include the possibility of describing
thermally or mechanically stable configurations under gravitational collapse.
However, a single three-dimensional boundary does not confine a finite
spacetime region of an eternal black hole. In order to describe black hole
thermodynamics starting from eternal black hole systems, it is necessary to
consider two three-dimensional timelike boundary surfaces $B_+$
and $B_-$ located in the  right $M_+$  and left $M_-$ regions of the spacetime.
This has been noted in Refs. \cite{FrMa2,FrMa1}, where the Hamiltonian,
quasilocal energy, and angular momentum for a finite region of a (distorted)
eternal black hole  have been constructed from the gravitational action.
In particular, the  Hamiltonian  for an eternal black hole is of the general
form  $H=H_+ - H_-$, where $H_+$ and $H_-$ are the Hamiltonian functions for
the
two separate wedges $M_+$ and $M_-$.

The aim of this paper is to generalize the
microcanonical functional integral (\ref{nu}) to  quantum self-gravitating
systems that include spacetimes whose topology and boundary conditions coincide
with the ones of (either distorted or Kerr-Newman) eternal black holes.
This  naturally requires the construction of the microcanonical
action (appropriate for fixed energy systems) when the  two
boundaries  $B_+$ and $B_-$  are located in the regions
$M_+$ and $M_-$  of an eternal black hole geometry.
The evaluation of the functional integral as well as its thermodynamical
consequences are discussed.
It turns out  that if the microcanonical sum over geometries for an eternal
black hole system is interpreted as its density of states, the  total entropy
of
the system  equals zero in the semiclassical approximation.  This result
applies
to the gravitational field itself of any type of eternal black holes (not only
of the Kerr-Newman form) for which the geometry is regular at the
bifurcation surface. Since in a microcanonical description it seems natural to
relate  the external and internal degrees of freedom of a black hole with the
boundary data at the surfaces $B_+$ and $B_-$ respectively \cite{FrMa2}, we
believe that the microcanonical functional integral for an eternal black hole
system opens the possibility of extending the path integral formulation of
gravitational thermodynamics to situations when  internal degrees of freedom
are present and allows the formulation of black hole
thermodynamics in terms of a single pure state.

The paper is organized as follows. We review in Section II the relevant
kinematical properties of a finite spacetime region  generated by the so-called
``tilted foliation" introduced in Ref. \cite{FrMa2} and compute its
microcanonical action.  The results are applied to the particular case of a
(distorted) Lorentzian eternal black hole. The microcanonical sum over
geometries  for a quantum gravitational system whose boundary conditions  equal
the boundary conditions of a physical Lorentzian, stationary, axisymmetric
eternal black hole is presented in Section III.  The path integral is
evaluated semiclassically by using  the  Lorentzian eternal black hole
metric that motivates its boundary conditions as well as a complex saddle point
of the microcanonical action.    The latter approximation  allows one to
understand the relationship between the functional integral for eternal black
holes and  the black hole density of states  computed in Ref. \cite{BrYo2}.
We
conclude in Section IV with general remarks concerning the construction of the
density of states for the ``exterior" region $M_+$   in terms of the
functional
integral for the complete spacetime, and the relevance of the results in a
thermofield dynamics interpretation of black hole thermodynamics.

\section{microcanonical action}

Consider a spacelike hypersurface $\Sigma$ with Einstein-Rosen
bridge topology $R^1 \times S^2$ whose  intrinsic geometry and  time
derivatives are chosen to satisfy the  gravitational constraint equations.
The evolution of these data  is
presumed to define a regular spacetime region to the future and past of the
 slice $\Sigma$ \cite{FrMa2}. We assume that
there exist two different spacelike hypersurfaces $\Sigma'$ and
$\Sigma''$ which intersect each other at a two-dimensional,
topologically spherical
spacelike  surface $S_0$.
The ``bifurcation" surface $S_0$  divides the slice $\Sigma$ in two parts
denoted by $\Sigma_+$ and $\Sigma_-$. The sequence of slices (generically
denoted by the symbol $\Sigma$ in what follows)  which intersect at the same
bifurcation surface $S_0$ is called a ``tilted foliation" \cite{FrMa2}. The
spacetime region $M$ lying between the two spacelike Cauchy surfaces  $\Sigma
'$
and $\Sigma ''$ consists therefore of two regions $M_+$ and $M_-$ (foliated by
$\Sigma_+$ and $\Sigma_-$ respectively) that join at $S_0$.  The  region $M$ we
consider is bounded not only by the slices $\Sigma'$ and $\Sigma''$ but also by
a three-dimensional timelike boundary $B$ that consists of two disconnected
parts $B_+$ and $B_-$.  For a general eternal black hole geometry the
boundaries   $B_+$ and $B_-$ are located in $M_+$ and $M_-$ respectively. The
intersections of the boundaries $B_+$ and $B_-$ with $\Sigma$ are topologically
spherical two-dimensional surfaces denoted by $S_+$ and $S_-$ respectively.
The topology of the  slices $\Sigma$ is therefore $I \times S^2$ (the interval
$I$ referring to a finite spatial distance), while the topology of the boundary
surfaces $B_\pm$ is $I \times S_\pm$ (the interval $I$ referring to a finite
time-like distance).

The line element of $M$ is of the general form \cite{FrMa2,ADM}
\begin{equation}
 ds^2 = -N^2 dt^2 + h_{ij} (dx^i + V^i dt)(dx^j + V^j dt)\ ,\label{metric}
\end{equation}
where   $N$ is the  corresponding lapse function and the spacelike
surfaces $\Sigma$ are chosen to coincide with surfaces of constant values of
$t$, so that the time coordinate $t$ is the scalar function that labels the
foliation. In particular, $\Sigma' = \Sigma_{t'}$ and $\Sigma'' =
\Sigma_{t''}$.
The
four-velocity vector $u^\mu$ is the timelike unit vector normal to the slices
$\Sigma$ and is defined by  $u_\mu = -N \, \partial_\mu t$.
Following \cite{BrYo1}, greek indices are used for tensors in $M$ while latin
indices are used for tensors defined in either $\Sigma$ of $B_\pm$.
The lapse function
$N$ is defined so that  $u \cdot u = -1$. The vector $t^\mu$ that connects
points with the same spatial coordinates is
\begin{equation}
t^\mu = Nu^\mu + V^\mu \ ,\label{tvector}
\end{equation}
so that $V^{i}= h^i _0 = - N u^i$ is the shift vector.
For the
``tilted foliations" considered here the slices corresponding to different
values of the parameter  $t$ join at the  bifurcation surface where
the lapse function $N$ vanishes.  The vector $u^\mu$ is chosen to be
future oriented in $M$ and the lapse $N$ is positive at
$\Sigma_+$ and negative at $\Sigma_-$.
The spacelike normal $n^\mu$ to
the three-dimensional boundaries $B_\pm$ is defined to be outward
pointing at
 $B_+$, inward pointing at $B_-$, and  normalized so that
$n \cdot n=+1$.  We shall assume that  the foliation  is further
restricted by the conditions $(u \cdot n)|_{B_{\pm}} = 0$ \cite{FrMa2,BrYo2}.

As argued in Ref. \cite{FrMa2}, it is convenient to define a set of ``standard"
coordinates $(t,x^i)$ for the ``tilted" foliation. These coordinates are in
a one-to-one correspondence with the
``standard" coordinates $(t, y, \theta, \phi)$ of a ``tilted"
foliation in a Schwarzschild-Kruskal spacetime.
The spatial coordinates $x^i$ have the same space orientation in
 both $R_+$ and $R_-$, but the time coordinate $t$ has opposite
orientations in $R_+$ and $R_-$.

The metric and extrinsic curvature of
$\Sigma$ as a surface embedded in $M$ are denoted by $h_{ij}$ and
$K_{ij} = -h_i ^{\ k} \nabla_{k}u_{j}$ respectively, while the metric and
extrinsic curvature of the boundaries $B_\pm$ as  surfaces embedded in $M$
are $\gamma_{ij}$ and  $\Theta_{ij} = -\gamma_i ^{\ k} \nabla_{k} n_{j}$
\cite{FrMa1,BrYo1}. Covariant differentiation  with respect to the metrics
$g_{{\mu\nu}}$ and $h_{ij}$ is denoted by $\nabla$ and $D$ respectively.
The induced metric and extrinsic curvature of the boundaries $S_\pm$ as
surfaces embedded on $\Sigma$ are  ${\sigma}_{ab}$ and  $k_{ab} =
-\sigma_{a}^{\ k}\, D_{k} n_{b}$ respectively $(a,b =2,3)$.  The normal vector
$n^{\mu}$ to ${B_\pm}$ is also the normal vector to $S_\pm$.
The extrinsic curvature tensors for the different surfaces are defined so that
\cite{BrYo1}
\begin{equation}
\Theta^{{\mu\nu}} = k^{{\mu\nu}} + u^\mu u^\nu n_\alpha a^\alpha + 2
\sigma^{\alpha ( \mu} u^{\nu )} n^\beta K_{\alpha \beta}\ ,
\label{Thetamunu}
\end{equation}
while the traces $\Theta$ and $k$ of the tensors $\Theta^{{\mu\nu}}$ and
$k^{{\mu\nu}}$ obey the relation
\begin{equation}
\Theta = k - n_\beta a^\beta \ ,\label{traces}
\end{equation}
where  the acceleration $a^\mu$ of the timelike
unit normal $u^{\mu}$ to the hypersurfaces $\Sigma$ is  $a^\mu =  u^\alpha
\nabla _{\alpha} u^\mu = (D^\mu N)/N$.
Finally, the determinants of the  metric tensors are related by
\begin{eqnarray}
\sqrt{-g} &=& |N| \sqrt{h}\ , \nonumber \\
\sqrt{-{\gamma}} &=& |N| \sqrt{\sigma} \ . \label{det}
\end{eqnarray}

As an illustration of a ``tilted" foliation, consider the simple case of
a static, spherically symmetric eternal black hole whose line element is
\cite{FrMa2}
\begin{equation}
ds^2 = -N^2 (y) dt^2 + dy^2 + r^2(y) d\Omega^2 \ .
\end{equation}
The set $(t, y, \theta, \phi)$ has the same spatial orientation
but differing time orientation in $R_+$ and $R_-$.
The coordinate $y$ represents the proper geodesic
distance from the ``throat"  of the Einstein-Rosen bridge at $S_0$.
The Hamiltonian constraint equation implies that
\begin{equation}
dy = \pm {{dr}\over {\sqrt{1 - {r_+}/r}}}
\end{equation}
in
$M_\pm$. It is convenient to choose $y$  positive in $\Sigma_+$,
negative in $\Sigma_-$, and zero at $S_0$
[so that $r(y=0) \equiv r_+$]. The solution is
regular at the surface $S_0$. The behavior of the gradient $r_{,y}$
exemplifies
an important property of eternal black holes:  the area of  two-dimensional
surfaces $S_+$ ($S_-$) in $\Sigma_+$ ($\Sigma_-$) increases (decreases) as
the proper coordinate $y$ increases. The lapse function in the
Schwarzschild-Kruskal spacetime is $N=\pm (1 - {r_+}/r)^{1/2}$ at $\Sigma_\pm$.
Observe that the gradient
$D_i N = N_{,y}\, {\delta}_{i}^{\ y}  = {r_+}/{2r^2} \,
{\delta}_{i}^{\ y}$, so that  $n^i D_i N = {r_+}/{2r^2}$ for both regions $M_+$
and $M_-$.

We turn now to consider the microcanonical action $S_m$ for a  ``tilted"
foliation.  The  action $S_m$ is the action appropriate to a variational
principle in which the fixed boundary conditions at the timelike  boundaries
$B_+$ and $B_-$ are not the spacetime three-geometry (that is, the metric
components $N$, $V^i$, and $\sigma_{ab}$) but the surface energy density
$\varepsilon$, surface momentum density $j_a$, and boundary metric
$\sigma_{ab}$ \cite{BrMaYo,BrYo1}.   The action  $S_m$ has been  constructed
for
spacetimes with a single timelike boundary in Refs. \cite{BrYo2,ensembles} by
adding the appropriate boundary terms to the ordinary gravitational action.
The surface energy density
$\varepsilon$ and  momentum density $j_a$ for a slice $\Sigma = \Sigma_+
\cup \Sigma_-$ of an eternal black hole spacetime has been calculated in
\cite{FrMa2} when the  two-dimensional boundary surfaces $S_+$
and $S_-$ are  located in either (1) the same space (either $\Sigma_+$ or
$\Sigma_-$), or (2) the two separate spaces $\Sigma_+$ and
$\Sigma_-$ respectively.
The  energy density $\varepsilon$ is the value (per unit boundary area) of the
Hamiltonian that generates unit time translations orthogonal to the boundaries
$S_+$ and $S_-$ \cite{BrYo2,FrMa2}. The surface momentum density $j_a$ is the
value (per unit boundary area) of the Hamiltonian that generates spatial
diffeomorphisms in the $\partial /{\partial x^i}$ direction on the
two-dimensional surfaces $S_+$ and $S_-$.  At each one of these surfaces the
energy  and  momentum densities are defined by
\begin{equation}
\varepsilon =  \bigl( k/\kappa\bigr) \label{edensity} \ ,
\, j_i = -2  \sigma_{ij} n_{k} P^{jk}/\sqrt{h}\ ,
\label{mdensity}
\end{equation}
where contributions due to functionals of the three-metrics at $B_+$ or
$B_-$  have been neglected.
The signs of the extrinsic curvatures $k$ of the surfaces $S_+$ and $S_-$
depend on the location of these surfaces for a chosen orientation of the normal
$n^\mu$. The quantities $\varepsilon$ and $j_i$, as well as their associated
integrated quantities, namely,  the quasilocal energy $E_\pm$ and angular
momentum $J_\pm$ for an eternal black hole, have been discussed in
\cite{FrMa2}.

The covariant form of the microcanonical action for a  general spacetime
$M$ generated by a ``tilted" foliation and whose respective
three-dimensional timelike surfaces $B_+$ and $B_-$ are located in $M_+$ and
$M_-$   can be written as
\begin{eqnarray}
S_{m} &=& {1\over2\kappa} \int_{M_+} d^4x \,\sqrt{-g}\, \Re  + {1\over\kappa}
\int_{({\scriptscriptstyle +})t'}^{t''} d^3x \,\sqrt{h}\, K  -{1\over\kappa}
\int_{{B_{\scriptscriptstyle +}}} d^3x  \,  \sqrt{-\gamma}  \,  t_{\mu}
{\Theta}^{{\mu\nu}} \,{\partial}_{\nu}t   \nonumber \\
&-&{1\over2\kappa} \int_{M_-} d^4x \,\sqrt{-g}\, \Re
+{1\over\kappa}   \int_{({\scriptscriptstyle -})t'}^{t''} d^3x \,\sqrt{h}\, K
-{1\over\kappa} \int_{{B_{\scriptscriptstyle -}}} d^3x  \,
\sqrt{-\gamma}  \,  t_{\mu} {\Theta}^{{\mu\nu}} \,{\partial}_{\nu}t \ ,
\label{Sm}
\end{eqnarray}
where  $\Re$ denotes the four-dimensional scalar curvature,
and $\kappa \equiv 8 \pi$. (We follow the conventions of
Ref. \cite{MTW} and units are chosen so that $G=\hbar=c=1$.)
The notation $\int_{({\scriptscriptstyle \pm})t'}^{t''}$  represents an
integral over the three-boundary $\Sigma_{\pm}$ at $t''$ minus an
integral over the three-boundary $\Sigma_{\pm}$ at $t'$.
The integrations are taken over
coordinates $x^\mu$ which possess the same orientation as the ``standard"
coordinates $(t, x^i)$ of the ``tilted" foliation. The differing signs in the
integrations over $M_+$ and $M_-$ reflect the fact that the coordinates have
different time orientations in $M_+$ and $M_-$.
The action
(\ref{Sm}) is independent of functionals of the three-metric at the
timelike boundaries $B_+$ and $B_-$ (``subtraction terms"), and reduces to the
microcanonical action introduced in Ref. \cite{BrYo2} when the spacetime region
is  bounded by a single timelike surface $B_+$.

The Hamiltonian form of the microcanonical action is easily obtained under a
$3+1$ spacetime split by recognizing that  there exists a direction of time at
the boundaries $B_+$ and $B_-$ inherited by the time vector $t^\mu$.
The four-dimensional scalar curvature is
\begin{equation}
\Re = R +  K^{\mu\nu} K_{\mu\nu} -
(K)^2 -2\nabla_\mu (K u^\mu + a^\mu)\ , \label{Re}
\end{equation}
where $R$ is the curvature scalar on $\Sigma$.
By using Gauss' theorem and  the conditions
\cite{BrYo2,FrMa2}
\begin{equation}
u \cdot n |_{B_\pm} = 0,\,  u \cdot a = 0,\,  u \cdot u = -1,\,n \cdot n = 1
\ ,
\end{equation}
as well as Eqns. (\ref{tvector}) and (\ref{Thetamunu}),
the action (\ref{Sm}) can be written as
\begin{eqnarray}
S_m &=& {1\over2\kappa} \int_{M_+} d^4x  \sqrt{-g}\,
\bigl[R + K_{\mu\nu} K^{\mu\nu} - (K)^2 \bigr]
+{1\over2\kappa}\int_{B_+} d^3x \sqrt{\sigma}\, n_i V_j (Kh^{ij} -
K^{ij}) \nonumber\\
&-& {1\over2\kappa} \int_{M_-} d^4x  \sqrt{-g}\,
\bigl[R + K_{\mu\nu} K^{\mu\nu} - (K)^2 \bigr]
-{1\over2\kappa}\int_{B_-} d^3x \sqrt{\sigma}\, n_i V_j (Kh^{ij} - K^{ij})
\ .
\label{Sm1}
\end{eqnarray}
In the most general case, there would be contributions to the action
(\ref{Sm1}) associated with the ``corners"  ${B''}_\pm =\Sigma'' \cap B_\pm$
and
${B'}_\pm = \Sigma' \cap B_\pm$, as well as with the cusp-like part $S_0$ of
the spacetime \cite{Ha,BrHa}. These contributions are related to the angles
between the  unit normal $u^\mu$ of $\Sigma$ and the spacelike normal $n^\mu$.
For simplicity, we consider here only the case when $u \cdot n =0$ at the
boudaries $B_\pm$.
 For a ``tilted" foliation  the contributions at $S_0$ connected with the
region
$M_+$ and $M_-$ have opposite signs and cancel identically due to the
regularity of the geometry at the bifurcation surface $S_0$
\cite{FrMa2}, and no extra contributions appear in (\ref{Sm1}).

The momentum $P^{ij}$ conjugate to the three-metric $h_{ij}$ of $\Sigma$
for the ``tilted" foliation can  be defined as  \cite{FrMa2}
\begin{equation}
P^{ij} = {1\over2\kappa} \sqrt{h}\, (K h^{ij} - K^{ij}) \ . \label{Pij}
\end{equation}
Since the sphere $S_0$ consists of  points which remain fixed under the
change of the parameter $t$, the time derivative of the
three-metric must vanish at $S_0$. The behaviour of the canonical variables in
the vicinity of the fixed sphere $S_0$  has been discussed in Ref.\cite{Ku76}.
Upon integration of the kinetic part of the volume integrals in
(\ref{Sm1})  the action becomes
\begin{equation}
 S_m = \int_M d^4x  \bigl[ P^{ij} \dot h_{ij} - N{\cal H} - V^i
{{\cal H}_i} \bigr]
\ , \label{HSm}
\end{equation}
where the dot denotes differentiation with respect to the global time $t$ and
the gravitational contribution to the Hamiltonian and momentum constraints are
given by the usual expressions
\begin{eqnarray} {\cal H} &=& (2\kappa)
G_{ijk\ell}\, P^{ij}\, P^{k\ell} - \sqrt{h}
      \, R/(2\kappa) \ , \nonumber \\
{{\cal H}_i} &=& -2 D_j \, P_i^{\ j} \ , \label{constraints}
\end{eqnarray}
with
$G_{ijk\ell} = (h_{ik} h_{j\ell} + h_{i\ell} h_{jk} - h_{ij} h_{k\ell})
/(2\sqrt{h}) $.

The microcanonical action  (\ref{HSm}) applies to any  smooth Lorentzian
geometry generated by a ``tilted" foliation when $B_+$ and $B_-$ are located
in the regions $M_+$ and $M_-$.  It has the same form as the ordinary canonical
action with no explicit boundary terms.   In particular, the
action (\ref{HSm})  vanishes identically for stationary solutions of  the
vacuum
Einstein equations describing stationary eternal black holes (with no extra
assumptions required about their symmetry).  In this case $\dot h_{ij}=0$, the
constraint equations are satisfied, and no boundary  terms remain in the
action.   This situation may of course  be different in the presence of matter
fields. For example, matter distributions at the horizon could  alter the
regularity of the geometry  there and give extra contributions to the action.

The ordinary gravitational  action $S$ for the
``tilted" foliation can be constructed from the microcanonical action
(\ref{HSm}) by adding boundary terms that change the boundary conditions from
fixed surface energy density $\varepsilon$, surface momentum density
$j_a$ and boundary metric $\sigma_{ab}$ at $B_\pm$ to fixed metric
components $N$, $V^i$, and $\sigma_{ab}$ at $B_\pm$ \cite{BrYo2,FrMa2}.  Two of
these boundary terms are needed.
The action $S$ is
\begin{eqnarray}
S &=& S_m - \int_ {B_{\scriptscriptstyle +}}
d^3x\,\sqrt{\sigma} \, \bigl[ N\varepsilon -V^ij_i \bigr]
+ \int_ {B_{\scriptscriptstyle -}} d^3x\,\sqrt{\sigma} \,
\bigl[  N\varepsilon -V^ij_i \bigr] \nonumber \\
&=& \int_M d^4x  \bigl[ P^{ij} \dot h_{ij} - N{\cal H} - V^i {{\cal H}_i}
\bigr]
 -\int_{B_+} d^3x \sqrt{\sigma} \bigl[ N\varepsilon - V^i j_i \bigr]
+\int_{B_-} d^3x \sqrt{\sigma} \bigl[ N\varepsilon - V^i j_i \bigr]
\ . \nonumber \\
\label{HS}
\end{eqnarray}
This form for the action $S$ and  its
consequences in the description of  eternal black holes have been discussed in
Ref. \cite{FrMa2}.

Consider finally,  as an illustration,  the microcanonical action for  a
spacetime region generated by the standard  ``untilted" foliation  when both
timelike boundaries $B_+$ and $B_-$ are located in the ``right" wedge $M_+$ of
an eternal black hole.  The foliation is regular everywhere in the region
between the initial $\Sigma'$ and final $\Sigma''$ slices. The global time
parameter $t$  labels the foliation and the
 four-velocity vector is   $u^\mu =-N {\delta^{\mu}}_{t}$, with the lapse
function being  positive everywhere in $M_+$ \cite{FrMa2}. In this case the
microcanonical action is
\begin{eqnarray}   S_{m} = {1\over2\kappa} \int_{M}
d^4x \,\sqrt{-g}\, \Re   +{1\over\kappa}  \int_{t'}^{t''} d^3x \,\sqrt{h}\, K
&-& {1\over\kappa} \int_{{B_{\scriptscriptstyle +}}}
d^3x  \, \sqrt{-\gamma}  \,  t_{\mu} {\Theta}^{{\mu\nu}} \,{\partial}_{\nu}t
\nonumber \\
&+& {1\over\kappa} \int_{{B_{\scriptscriptstyle -}}} d^3x  \,
\sqrt{-\gamma}  \,  t_{\mu} {\Theta}^{{\mu\nu}} \,{\partial}_{\nu}t \ .
\end{eqnarray}
It is easy to show that the Hamiltonian version of this action is also given by
Eqn. (\ref{HSm}).
The difference between the microcanonical action
for ``tilted" and ``untilted"
foliations manifests itself in their boundary data.
(For instance, since the sign of the surface energy density
$\varepsilon_-$ is connected with the sign of extrinsic curvature of
the surface $S_-$ for a chosen orientation of the normal $n^\mu$, the sign
 of $\varepsilon_-$  when $S_-$ is
 located in $M_+$ for the ``untilted" foliation is opposite to the sign of
$\varepsilon_-$  when $S_-$ is located in $M_-$ for the ``tilted" foliation.)
The Hamiltonian form of the microcanonical action for ``untilted" foliations
has been used in Refs. \cite{BrYoRev,Te}  when the two
three-dimensional
 boundaries of the spacetime are located in the single complex
sector of an ordinary black hole and the internal boundary approaches the black
hole horizon.  We would like to emphasize that, even if the
microcanonical actions for ``tilted" and ``untilted" foliations reduce to
similar Hamiltonian forms, the former applies  to spacetimes whose two regions
intersect at a fixed surface $S_0$. The action (\ref{HSm})  is the
necessary action to describe the dynamics of finite regions of a  distorted
eternal black hole and will play an important role in the sum over geometries
for eternal black hole systems presented below.

\section{functional integral}

We consider in this section a microcanonical functional integral
for a physical
system whose boundary conditions correspond to the ones of an eternal version
of a  black hole. Consider first the functional integral for a microcanonical
gravitational system for which two timelike boundary
surfaces $B_+$ and $B_-$
are needed in order to contain a finite spacetime region.
The functional integral takes the form
\begin{equation}
{\bar {\nu}}[\varepsilon_+, j_+, \sigma_+ ;
\varepsilon_-, j_-, \sigma_-] = \sum_{\sl M} \int {\cal D}H \exp(i S_m) \ ,
\label{ournu}
\end{equation}
and is a  functional of the  energy density $\varepsilon$,
 momentum density $j_a$, and two-metric $\sigma_{ab}$ at the
boundaries $B_+$ and $B_-$. For simplicity the notation $j_\pm$ indicates that
the quantity $j_a$ is specified at the surface $B_\pm$.
The sum over $M$  refers to a sum over manifolds of different
topologies whose boundaries have topologies
$B_+ = S_+ \times S^1 = S^2 \times S^1$ and
$B_- = S_- \times S^1 = S^2 \times S^1$. The element $S^1$ is due to the
periodic identification in the global time direction at the boundaries when the
initial and final hypersurfaces are identified.   The  integral is a  sum over
periodic Lorentzian metrics that satisfy the boundary conditions at $B_+$ and
$B_-$. The action appearing in (\ref{ournu}) is the microcanonical action
$S_m$ discussed  in Section II, but with the boundary terms corresponding
to $\Sigma'$ and $\Sigma''$ dropped because the manifolds summed over possess
only two boundary elements, namely, $B_+$ and $B_-$.

As with the density of states (\ref{nu}) \cite{BrYoRev}, the functional
integral (\ref{ournu}) can be considered as the result of tracing over
initial and final configurations in a
microcanonical density matrix $\rho_m$ of the form:
\begin{equation}
{\bar{\nu}}[\varepsilon_+, j_+, \sigma_+ ; \varepsilon_-, j_-, \sigma_-] =
\int {\cal D}h \, \rho_m[h, h; {\alpha''}_{\pm}, {\alpha'}_{\pm};
\varepsilon_+,
j_+,  \sigma_+ ; \varepsilon_-, j_-, \sigma_-] \ ,\label{rho}
\end{equation}
where the angles ${\alpha''}_{\pm}$ and ${\alpha'}_{\pm}$ at the corners
${B''}_\pm$ and ${B'}_\pm$ are required to satisfy the condition
${\alpha''}_{\pm} + {\alpha'}_{\pm} = \pi$ to guarantee the smoothness of the
boundaries $B_+$ and $B_-$.

Consider now the functional integral (\ref{ournu})
in the case when the boundary
surfaces $B_+$ and $B_-$ are located in separate regions $M_+$ and $M_-$ and
the
fixed boundary data $(\varepsilon_+, j_+, \sigma_+)$  and  $(\varepsilon_-,
j_-,
\sigma_-)$ correspond to the boundary data of a general Lorentzian, stationary,
axisymmetric eternal black hole.  This spacetime is a solution of Einstein
equations whose line element is of the form (\ref{metric}):
\begin{equation}
ds^2 = -{\tilde N}^2 dt^2 + {\tilde h}_{ij} (dx^i + {\tilde V}^i dt)
(dx^j + {\tilde V}^j dt)\ , \label{LBH}
\end{equation}
where
the lapse ${\tilde N}$, shift vector ${\tilde V}^i$, and three-metric
${\tilde h}_{ij}$ are particular functions of the  spatial coordinates $x^i (i=
1,2,3)$. For convenience, the spatial coordinates can be chosen  to be
co-rotating with the horizon \cite{Ba,BrMaYo}, so that  ${\tilde V}^i/{\tilde
N}=0$ at the horizon. In this spacetime the spacelike slices $\Sigma$ are
constant stationary time surfaces that contain the closed orbits of the axial
Killing vector field.  The two-dimensional boundaries $S_+$ and $S_-$ of
$\Sigma$ also contain the orbits of the axial Killing field. The boundary
data $(\varepsilon_+, j_+, \sigma_+)$  and  $(\varepsilon_-, j_-, \sigma_-)$ of
this solution can be determined at $S_+$ and $S_-$ for each  slice $\Sigma$.
By virtue of the gravitational constraint equations, these data determine
uniquely the size of the black hole horizon \cite{ensembles} and are such that
the two-metric ${\tilde \sigma}_{ab}$ is continuous at this horizon.  We will
assume that both boundaries $S_+$ and $S_-$ of the rotating solution  used to
generate the boundary data are not located beyond the speed-of-light surfaces
surrounding the black hole \cite{BrMaYo,FrTh}. The eternal black hole
functional
integral ${\bar\nu}_{*}$ is given by expression (\ref{ournu}) when the boundary
data at  $B_+$ and $B_-$ of the geometries summed over  coincide with the  data
of the classical Lorentzian eternal black hole.
 The topology of each one of these
spacetimes is arbitrary but each boundary $B_\pm$ is required to have the
boundary topology $S_\pm \times S^1$.

We evaluate now the functional integral in the semiclassical
approximation. This requires finding a four-metric that extremizes the
action $S_m$ and satisfies the boundary conditions  $(\varepsilon_+, j_+,
\sigma_+)$  at  $S_+$ and  $(\varepsilon_-, j_-, \sigma_-)$
at $S_-$. Fortunately, the
Lorentzian eternal black hole metric
(\ref{LBH})   can be
periodically identified in the global time direction and placed on a
manifold  whose two spatial boundaries have the desired topologies
$S_\pm \times S^1$.  The periodic identification alters neither the constraint
equations nor the boundary data and the resulting metric can be used to
approximate the path integral.
As observed in \cite{BrYo2}, if the
physical system can be approximated by a single classical configuration, this
configuration will be the real spacetime (\ref{LBH}) that induced the
boundary data.
In the semiclassical approximation the functional integral ${\bar{\nu}}_{*}$
becomes
\begin{equation}
{\bar{\nu}}_{*}[\varepsilon_+,j_+, \sigma_+;  \varepsilon_- ,j_-,\sigma_-]
\approx   \exp \big( i S_m[{\tilde N}, {\tilde V}, {\tilde h}] \big)\ ,
\label{seminuL} \end{equation}
where the action $S_m[{\tilde N}, {\tilde V}, {\tilde h}]$ is the
microcanonical action evaluated at the periodic manifold (\ref{LBH}).

The action $S_m[{\tilde N}, {\tilde V}, {\tilde h}]$ is obtained from
 (\ref{Sm}) by dropping the integrals at $t'$ and $t''$, and  its Hamiltonian
form is given by Eqn. (\ref{HSm}).  This action
vanishes identically: the volume  term equals zero because $P^{ij} \dot h_{ij}$
is zero by stationarity and the gravitational constraints are satisfied. The
functional integral is  therefore
\begin{equation}
{\bar{\nu}}_{*}[\varepsilon_+ ,j_+, \sigma_+; \varepsilon_-,
j_-, \sigma_-]  \approx \exp \big( 0 \big) = 1 \, \label{nufinal}
\end{equation}
in the semiclassical approximation.

It is illustrative to consider now a
complex  four-metric which also extremizes the microcanonical action  for
eternal black hole  boundary conditions and which can be used to reevaluate the
path integral (\ref{ournu}) in a steepest descent approximation.
This alternative approximation of the quantity ${\bar{\nu}}_{*}$ is useful in
understanding the relationship of the result (\ref{nufinal}) with the density
of states for an ordinary  (that is, non-eternal) black hole computed in Ref.
\cite{BrYo2}. The
complex metric can be obtained from the
Lorentzian eternal black hole metric (\ref{LBH}) by replacing the stationary
time $t$ with imaginary time, namely, $t  \to -it$, with $t$ real. Its line
element is
\begin{equation}
ds^2 = -(-i{\tilde N})^2 dt^2 + {\tilde h}_{ij}  (dx^i
-i{\tilde V}^i dt)(dx^j -i{\tilde V}^j dt) \ ,\label{cmetric}
\end{equation}
with ${\tilde N}$, ${\tilde V}^i$, and ${\tilde h}_{ij}$ real.
The complex metric has $(-i{\tilde N})$ as its lapse function and $(-i{\tilde
V}^i)$ as its shift vector, with ${\tilde N}$ being real and positive in $M_+$
and real and negative in $M_-$. (The metric becomes Euclidean if ${\tilde V}^i
=0$.)
The complexification  map $\Psi$ defined by  $\Psi(N)= -iN$,
$\Psi(V^i) = -iV^i$ is equivalent to transforming  the global vector $t^\mu$ so
that  $t^\mu \to \exp(i\vartheta ) t^\mu$, with $\vartheta = -\pi /2$
\cite{BrMaYo}. In particular, $\Psi(|N|) = -i|N|$. Under the map $\Psi$ and the
periodic identification in the time-like direction, the ``right" and ``left"
wedges of a Lorentzian eternal black hole are mapped into two complex sectors
(which we denote  ${\bar M}_+$ and ${\bar M}_-$ for simplicity).
\footnote{The complexification $\Psi$ maps the ``right" and ``left" Lorentzian
wedges of an eternal black hole  into distinct complex sectors. This can be
seen  by considering a finite matter distribution located at a finite distance
in one of the regions of a  static Lorentzian black hole. Because of the
presence of matter, the complexification $\Psi$ produces two complex sectors
that cannot be identified.}

The complexification map $\Psi$  preserves the reflection symmetry and the
canonical variables $h_{ij}$ and $P^{ij}$ of the Lorentzian eternal black hole
solution.  This implies that the microcanonical boundary data (constructed
uniquely from those canonical variables) that characterize the
real Lorentzian solution and the functional integral are also the
boundary data of the complex metric (\ref{cmetric}).
As  pointed out  in Ref. \cite{BrYoRev}, this property guarantees  that the sum
over geometries extremized by the complex eternal black hole metric will indeed
describe the physical properties of a real Lorentzian eternal black hole in the
semiclassical approximation.   The  complexification map $\Psi$ is in fact the
only complexification map that preserves the boundary data of the Lorentzian
solution.  Complexifications of the type $N \to -i N$ for ${\bar M}_+$ and
$N \to iN$ for ${\bar M}_-$ would produce complex metrics whose boundary
surface energy densities do not coincide with the boundary surface energy
densities of the Lorentzian eternal black hole. This can be checked by  using
the explicit expressions presented
in \cite{FrMa2} for the quasilocal energy  of the latter solution.

The complex geometry consists of two complex sectors ${\bar M}_+$ and
${\bar M}_-$ which join at the locus of points at which ${\tilde N}=0$.
 For each sector the two-surface at which ${\tilde N}=0$ is called a
``bolt" \cite{GiHa2}. The geometric structure of each of these sectors
resembles the structure of  the   single black hole complex sector used in
Refs.
\cite{BrMaYo,BrYo2} to approximate black hole functional integrals.
Since the Lorentzian metric is a solution of Einstein equations, the
complex metric (\ref{cmetric}) is also a solution of Einstein equations with
the
exception of the locus ${\tilde N} =0$.
Einstein equations are not satisfied at the ``bolt" if a conical
singularity exists there for every $\Sigma$.
Each sector ${\bar M}_+$ and ${\bar M}_-$ has consequently the topology of a
``punctured" disk $\times S^2$  because the two-space defined by the
plane generated by the unit normals $u^\mu$ and $n^\mu$ has the topology of a
``punctured" disk \cite{BrYo2}.
 The outer three-dimensional boundaries of the
sectors ${\bar M}_+$ and ${\bar M}_-$ are  $B_+$ and $B_-$, while
their inner three-dimensional boundaries are  denoted by $^3\! H_+$ and  $^3\!
H_-$ respectively.
The boundary data $(\varepsilon_+, j_+, \sigma_+)$ and $(\varepsilon_-, j_-,
\sigma_-)$ are specified at $B_+$ and $B_-$.
The outer boundaries $B_\pm$ of $M_\pm$ have topologies $S_\pm \times S^1$
while the
inner boundaries $^3\! H_\pm$ of $M_\pm$ have topologies
$^2\! H_\pm \times S^1$,
 where $^2\! H_+$ and $^2\! H_-$ denote respectively the
intersection of the slices $\Sigma_+$ and $\Sigma_-$ with the black hole
horizon
for the Lorentzian metric.
Each one of the slices $\Sigma_+$ and $\Sigma_-$ of the complex metric has the
topology $I \times S^2$ due to the openings at $^3\! H_+$ and  $^3\! H_-$.

To satisfy the vacuum Einstein
equations and assure the smoothness of the complex geometry it is necessary
to impose regularity conditions in the submanifolds that contain the
unit normals $n^i$ to the ``bolt" for each surface $t={\rm const.}$
\cite{BrMaYo,BrYo2}
and to require the two-metric $\sigma_{ab}$ to be
continuous at $^2\! H_+$ and  $^2\! H_-$.
As one approaches the ``bolt"  from both ${\bar M}_+$ and ${\bar M}_-$ the
metric becomes Euclidean
\begin{equation}
ds^2 \approx {\tilde N}^2 dt^2+ {\tilde h}_{ij} dx^i dx^j \ .
\end{equation}
The regularity  is enforced if, for each sector ${\bar M}_+$ and
${\bar M}_-$, the proper
 circumference of circles surrounding the ``bolt" equals
$2\pi$ times their proper distance to the ``bolt". The proper circumference is
given by $P |{\tilde N}|$ in both ${\bar M}_+$ and ${\bar M}_-$, where $P$
denotes the period of the geometry in coordinate (stationary) time $t$.  The
complexification map $\Psi$ guarantees  that  the unit normals ${\tilde n}^i$
to
the ``bolt" for each surface $\Sigma$ are continuously defined.
 Because of this,  the regularity conditions at the ``bolt" as approached from
either region ${\bar M}_+$ and ${\bar M}_-$ take the form
\begin{equation}
P ={{ 2\pi} \over{ {\tilde n}^i
D_i{\tilde N}} } \ .
\label{regcon}
\end{equation}
As mentioned in Section II, the quantity  ${\tilde n}^i D_i {\tilde N}$
(defined in terms of the ``standard" coordinates) has the same  relative signs
in both regions ${\bar M}_+$ and ${\bar M}_-$ of an eternal black hole.
Condition (\ref{regcon}) holds at each point on the bolt \cite{BrYo2}.

The regularity conditions (\ref{regcon})  and
the requirement that  ${\tilde N} =0$ at  $^3\! H_+$ and  $^3\!H_-$ assure the
smoothness of the complex geometries by  sealing the openings at
$^3\! H_+$ and  $^3\! H_-$ with no conical singularities. They effectively
guarantee the absence of inner boundaries for either sector ${\bar M}_+$
or ${\bar M}_-$ and imply that  the plane generated by the normals $u^\mu$ and
$n^\mu$
 becomes a smooth disk with $R^2$ topology.  The topology of each sector
${\bar M}_+$ and ${\bar M}_-$ becomes $R^2 \times S^2$.
In this way the conditions mentioned above amount  to the absense of
inner boundary information \cite{BrYoRev} at either $^3\! H_+$ or $^3\! H_-$.
However, each element $^3\! H_+$ and $^3\! H_-$ does contribute a term to the
microcanonical action for the complex geometry (\ref{cmetric}). For an ordinary
black hole the contribution  from the single inner element  $^3\! H_+$ to the
action is  indeed responsible for the black hole entropy. In the present case,
two such contributions to the action arise at $^3\! H_+$ and $^3\! H_-$, and it
becomes important to determine whether they  either add or
cancel each other.

The complex metric periodically identified with a coordinate period satisfying
(\ref{regcon}) is an extremum of the action $S_m$
and satisfies the desired boundary conditions.
It is not included in the sum over Lorentzian
geometries ${\bar{\nu}}_{*}$ in (\ref{ournu}) but it can be used to
approximate it by distorting the contours of
integration for both the lapse ${\tilde N}$ and the shift ${\tilde V}^i$
into the complex plane \cite{BrYo2}. In this approximation the functional
integral becomes
\begin{equation}
{\bar{\nu}}_{*}[\varepsilon_+,j_+, \sigma_+;  \varepsilon_- ,j_-,\sigma_-]
\approx
\exp \big( i S_m[-i{\tilde N}, -i{\tilde V}, \tilde h]\big)\ . \label{seminu}
\end{equation}
The  action $S_m[-i{\tilde N}, -i{\tilde V}, {\tilde h}]$ is the microcanonical
action  (\ref{Sm})  for a ``tilted" foliation
evaluated at the complex metric (\ref{cmetric}) when the smoothness of the
geometries at  $^3 \! H_+$ and $^3 \! H_-$ is inforced. As before, no integrals
at $t'$ and $t''$ are present because of the periodic identification of
$\Sigma'$ and $\Sigma''$ in the complex manifold.  However, we cannot use Eqn.
(\ref{HSm}) directly to evaluate the action for the complex metric since the
latter action must include terms at both $^3 \! H_+$ and $^3 \! H_-$. If one
repeats the Hamiltonian decomposition  (\ref{Re})-(\ref{Pij})  of the
Lorentzian
action  when two internal ``boundary" elements exist at $^3 \! H_+$ and
$^3 \! H_-$ one obtains
\begin{eqnarray}
 S_m = \int_M d^4x  \bigl[ P^{ij} \dot h_{ij} - N{\cal H} - V^i
{{\cal H}_i} \bigr]
&+& \int_{H^+} d^3x \sqrt{\sigma}\,
\bigl( |N| n^\mu a_\mu  /\kappa +  2 n_i V_j P^{ij} /\sqrt{h} \bigr)
\nonumber \\
&+& \int_{H^-} d^3x \sqrt{\sigma}\,
\bigl( |N| n^\mu a_\mu  /\kappa -  2 n_i V_j P^{ij} /\sqrt{h} \bigr)
\ . \label{HSm+b}
\end{eqnarray}
This action can now be used to evaluate the action
$S_m[-i{\tilde N}, -i{\tilde V}, {\tilde h}]$. The volume term in the latter
action vanishes  due to stationarity  and to the
Hamiltonian and momentum constraints being satisfied by the complex metric
(\ref{cmetric}). Since the shift vector ${\tilde V}^i$ vanishes at  both
$^3 \! H_+$ and
$^3 \! H_-$, only the terms involving the acceleration of the unit
normal $u^\mu$ remain to be evaluated. By using the regularity conditions
(\ref{regcon}) and  the expression
 ${\tilde a}_i = (D_i {\tilde N})/{\tilde N}$, the action at the complex
metric  becomes
\begin{eqnarray}
S_m[-i{\tilde N}, -i{\tilde V}, {\tilde h}] &=&
-{i \over \kappa} \int_{^3\! H_{\scriptscriptstyle +}} d^3x\sqrt{\tilde
\sigma}  \,|{\tilde N}|\, {\tilde n}^\mu \, {\tilde a}_\mu
-{i \over \kappa} \int_{^3\! H_{\scriptscriptstyle -}} d^3x\sqrt{\tilde
\sigma}  \,|{\tilde N}|\, {\tilde n}^\mu \,{\tilde a}_\mu \nonumber \\
&=& -{i \over \kappa} \int_{^2\! H_{\scriptscriptstyle +}} d^2x
\sqrt{\tilde \sigma} \,P  \, {\tilde n}^i \, D_i{\tilde N}
+{i \over \kappa} \int_{^2\! H_{\scriptscriptstyle -}} d^2x \sqrt{\tilde
\sigma}\, P \, {\tilde n}^i \, D_i {\tilde N} \nonumber \\
&=& -{{2\pi i}\over \kappa}\int_{^2\! H_{\scriptscriptstyle +}} d^2x
\sqrt{\tilde \sigma}\,
 +{{2\pi i}\over \kappa}\int_{^2\! H_{\scriptscriptstyle -}} d^2x
\sqrt{\tilde \sigma} \nonumber \\
 &=& -{i \over 4} A_+
+{i \over 4} A_- \ ,
\end{eqnarray}
where $A_+$ and $A_-$ denote the surface area of the horizon elements
${^2\!H_{\scriptscriptstyle +}}$ and  ${^2\! H_{\scriptscriptstyle -}}$.
The gravitational constraint
 equations imply that $A_+$ and $A_-$ are functions of the boundary data
$(\varepsilon_+, j_+, \sigma_+)$ and $(\varepsilon_-, j_-, \sigma_-)$
respectively.
The functional integral (\ref{seminu})
is therefore
\begin{equation}
{\bar{\nu}}_{*}[\varepsilon_+, j_+, \sigma_+; \varepsilon_-, j_-, \sigma_-]
\approx
\exp{\Bigg({1\over 4}A_+ - {1\over 4}A_- \Bigg)} \ . \label{subtra}
\end{equation}
Recall that the data $(\varepsilon_+, j_+, \sigma_+)$ and
$(\varepsilon_-, j_-, \sigma_-)$ correspond to the boundary data of the
classical Lorentzian eternal black hole solution (\ref{LBH}). As such, they are
not an arbitrary set of boundary data but a set that guarantees that the
two-metric is continuous at the horizon of the Lorentzian black hole. Since the
periodic identification and the complexification $\Psi$ do not alter these
boundary data nor the gravitational constraint equations, the area $A_+$ of
${^2\!H_{\scriptscriptstyle +}}$ coincides with the area $A_-$ of
${^2\!H_{\scriptscriptstyle -}}$:
$A_+ (\varepsilon_+, j_+, \sigma_+) = A_- (\varepsilon_-, j_-, \sigma_-) \equiv
A_H$.
This implies that, in agreement with (\ref{nufinal}), the eternal black hole
functional integral is
\begin{equation}
{\bar{\nu}}_{*}[\varepsilon_+ ,j_+, \sigma_+; \varepsilon_-,
j_-, \sigma_-]  \approx \exp{\Bigg( {1\over 4}A_H - {1\over 4}A_H \Bigg)} = 1
\,
\label{nufinal2}
\end{equation}
in the ``zero-loop" approximation.

If the microcanonical functional integral (\ref{ournu})  is
interpreted as the density of states of the statistical system, it is possible
to express ${\bar{\nu}}_{*}$ approximately  as
\begin{equation}
{\bar{\nu}}_{*}[\varepsilon_+, j_+, \sigma_+; \varepsilon_-, j_-, \sigma_-]
\approx \exp({\cal S}
[\varepsilon_+, j_+, \sigma_+; \varepsilon_-, j_-, \sigma_-]) \ ,
\end{equation}
where ${\cal S}$ represents the total entropy of the
system.  The result  (\ref{nufinal2}) implies that  the entropy  for the system
is
\begin{equation}
{\cal S} \approx  {1\over 4}{A_H} - {1\over 4} {A_H}  = 0
\label{entropy}
\end{equation}
in the semiclassical  approximation.
Notice that the total entropy is given formally by the subtraction
${\cal S} = {{\cal S}_+}[\varepsilon_+ ,j_+, \sigma_+]  - {{\cal
S}_-}[\varepsilon_- ,j_-, \sigma_-]$, where both ${{\cal S}_+}$ and
${{\cal S}_-}$ equal one fourth of the area of the horizon in this
approximation
and can be interpreted as the semiclassical entropies associated with the
external ($M_+$) and internal ($M_-$) regions respectively of the eternal black
hole system.

\section{Concluding remarks}

The functional integral (\ref{seminuL}) and (\ref{nufinal2}) refers to a
quantum-statistical system  which is classically approximated by a general
stationary, axisymmetric, eternal black hole solution of Einstein equations
within a  region bounded by two timelike  surfaces $B_+$ and $B_-$.
If the functional integral is interpreted as the density of states of the
system,  the entropy of the latter in the semiclassical approximation equals
${\cal S} = A_H/4 - A_H/4 = 0$, where $A_H$ is the area of the horizon of the
physical eternal black hole solution that classically approximates the system.
This result is a consequence of the choice of boundary data, the gravitational
constraint equations, and the vanishing of the microcanonical action for the
four-geometries that satisfy  the boundary conditions and approximate the path
integral.

Although the result  (\ref{entropy}) for the entropy can be expected on
physical grounds, it is important to stress its generality.
Since the spacetime is not  necessarily asymptotically flat outside the
boundaries $B_+$ and $B_-$, the physical eternal black hole that
approximates the quantum system is in general a distorted black hole not
necessarily of the Kerr-Newman form.
Expression (\ref{entropy}) applies to any of these configurations
 in the strong gravity regime
 (even in the case when  gravitational perturbations are not small)
 since the functional
 integral refers to the gravitational field itself of any type
of spacetime whose geometry is
regular at the bifurcation surface and which satisfies eternal black hole
boundary conditions. As is the case for the ordinary black hole entropy
computed
in \cite{BrYo2}, the entropy  (\ref{entropy}) does not seem to depend on
axisymmetry. These results indicate  that a pure state (of zero
entropy) can be defined not only for
matter fields perturbations propagating in the spacetime of an eternal black
hole but also for the gravitational field itself. This is
physically appealing: the initial data for the eternal black hole specified at
the spacelike hypersurface $\Sigma$  contain all the information required for
the evolution of both the exterior and interior parts of a physical black hole.
The entropy associated with $\Sigma$ must therefore equal zero.

These conclusions are in complete agreement with thermofield dynamics
descriptions of quantum processes \cite{UmTa} and, in particular, with the
application of this approach to black hole thermodynamics developed originally
by Israel \cite{Is} for small perturbations (see also Refs. \cite{tfd}). In the
original formulation of thermofield dynamics an extended Fock space ${\cal F}
\otimes {\tilde {\cal F}}$  is obtained by augmenting the physical Fock space
${\cal F}$ by a ``fictitious" Fock space ${\tilde {\cal F}}$. A pure vacuum
state in the extended Fock space ${\cal F} \otimes {\tilde {\cal F}}$
corresponds to a mixed state in the physical Fock space ${\cal F}$.  In the
application of this approach to black hole processes the Boulware states of
particles
in the two causally disconnected regions  $R_+$ and $R_-$ of an eternal
black hole  can be identified with the spaces  ${\cal F}$ and ${\tilde {\cal
F}}$ respectively, and the  space ${\cal F} \otimes {\tilde {\cal F}}$
describes
states for the complete system. The results of  Ref. \cite{FrMa2}
regarding the gravitational Hamiltonian $H = H_+ - H_-$ for a spatially bounded
region of an eternal black hole and the thermodynamical functional
integral for eternal black holes presented in this paper strongly indicate that
the thermofield dynamics description of quantum field processes in a curved
background can be extended beyond small perturbations to the gravitational
field
itself of distorted eternal black holes.

The microcanonical functional integral (\ref{seminu})  reflects the
properties of a pure state of zero entropy. It would be specially interesting
to recover the density of states and entropy for ``mixed" states in the
``exterior region" $M_+$ of an eternal black hole  by explicitly tracing out
in (\ref{seminu}) the  internal degrees of freedom of the  black hole itself.
This operation  must yield the  density of states $\nu_{*}$ for a black hole
computed in \cite{BrYo2} (with a corresponding entropy given by one fourth of
the horizon area) in the semiclassical approximation. It is not yet clear how
to
perform this ``tracing" operation satisfactorily beginning with the functional
integral (\ref{seminu}). There are several ways in which one could proceed.
For
example, it has been suggested \cite{FrMa2} that the internal degrees of
freedom
of a black hole can be identified with the set of  boundary data specified at
the boundary $B_-$.
 One could formally construct a functional integral $\nu_{*}$
on $M_+$ by integrating over these boundary data in the form
\begin{equation}
\nu_{*}[\varepsilon_+, j_+, \sigma_+] \approx \int
{\cal D}\mu[\varepsilon_-, j_-, \sigma_-]  \,\,
{\bar{\nu}}_{*}[\varepsilon_+, j_+, \sigma_+; \varepsilon_-, j_-, \sigma_-] \ ,
\label{trace}
\end{equation}
where ${\cal D}\mu[\varepsilon_-, j_-, \sigma_-]$ denotes some measure in the
space of boundary data at $B_-$. The definition of this measure is delicate.
Since the initial data $(\varepsilon_-, j_-,\sigma_-)$ at $B_-$ uniquely
determine the horizon area $A_-$ in a  microcanonical description (see, for
example, Ref. \cite{ensembles}), the measure ${\cal D}\mu$ may be tentatively
regarded as  proportional to the differential $d{\cal S}_-$ of the entropy
${\cal S}_-$ in a first approximation. (Although in a different context, this
measure has been previously considered in  Ref. \cite{WhYo}.) If (\ref{subtra})
is substituted directly  in (\ref{trace}) the integral would become
\begin{equation}
\nu_{*}[\varepsilon_+, j_+, \sigma_+]
 \approx \int _0 ^{\infty} d{\cal S}_- \,\,
\exp ({\cal S}_+ -{\cal S}_-) \approx \exp ({\cal S}_+)  \ . \label{lastnu}
\end{equation}
While this is the desired result for the quantity $\nu_{*}$, the approach has
several obvious conceptual difficulties. For a given value of $A_+$, the
integration (\ref{lastnu}) implies a sum over the whole range of areas $A_-$.
This ``decoupling" of the ``degrees of freedom" $A_+$ and $A_-$ is not a
semiclassical effect because the boundary data $(\varepsilon_\pm, j_\pm,
\sigma_\pm)$ at $B_\pm$ for a  classical eternal black hole are such that
$A_+ = A_- = A_H$ in the absence of matter at the horizon. The integral
(\ref{lastnu}) therefore represents a sum over quantum spacetimes which satisfy
the boundary data at $B_\pm$ but whose two-metric is not regular at the
bifurcation surface $S_0$. However, it is not clear whether the expression
${\bar{\nu}}_{*} \approx \exp({\cal S}_+ - {\cal S}_-)$ is appropriate for
non-smooth geometries. The contribution of these geometries to the functional
integral (\ref{ournu}) could perhaps be calculated using the approaches
developed in Refs. \cite{Ha,BrHa,HaLo}.

Another approach to recover the black hole density of states $\nu_{*}$ from the
eternal black hole functional integral ${\bar{\nu}}_{*}$ is the following. The
quantity $\nu_{*}$ computed in \cite{BrYoRev} is obtained as the trace of a
density matrix when a special set of conditions (which include ${\tilde N}=0$,
${\tilde V}^i =0$, and the regularity conditions) is imposed at the ``bolt" of
a complex geometry. These conditions imply that the complex sector has no inner
boundary. Similarly, the eternal black hole functional integral
${\bar{\nu}}_{*}$ computed in Section III is obtained as the trace (\ref{rho})
of a density matrix when similar conditions are imposed at  $^3\! H_+$ and
$^3\!
H_-$. However, it is not difficult to see that  ${\bar{\nu}}_{*}$ would equal
$\nu_{*}$ if the above conditions are only imposed at $^3\! H_+$ while
microcanonical boundary conditions are imposed  at $^3\! H_-$. Tracing out
internal degrees of freedom would seem to be equivalent to imposing
microcanonical boundary conditions at  $^3\! H_-$ in the functional integral
(\ref{ournu}). If this procedure is physically sensible, the geometries summed
over in the tracing operation will not be smooth at the ``bolt". It might be
interesting to study the relationship between this approach and the proposals
for black hole entropy presented in Refs. \cite{CaTe,Te}, and to reproduce
the thermodynamical results presented in this paper by using the Hamiltonian
methods developed in Refs. \cite{LoWh,Ku1,FrMa2}.

Finally, the relationship between vacuum states in the left and right
wedges of the Kruskal diagram to the Hartle-Hawking vacuum for quantum
fields defined on the maximally extended black hole is well known
\cite{Is,HaHa,Ja,BaFrZe}. Recently,  the Hartle-Hawking vacuum state for
linearized field perturbations for all fields has been constructed by using a
no-boundary wave function proposal for a black hole \cite{BaFrZe}.   The
essential properties defining a general Hartle-Hawking state have been
described
in Ref. \cite{Ja}.    It would be interesting to understand the
significance of the thermodynamical functional integral presented here in the
construction of the  Hartle-Hawking vacuum state (within properly defined
boundary surfaces that do not exceed the speed-of-light surfaces) for
stationary, axisymmetric black holes in the strong gravity regime when the
perturbations of the gravitational field are not necessarily small.

\acknowledgments

It is a pleasure to thank Valeri Frolov for his inspiration and for many
stimulating discussions. The author is also indebted to Werner Israel for his
encouragement and for his critical remarks, and to Andrei
Barvinsky, Geoff Hayward, and Andrei Zelnikov for useful conversations.
Research support was received from the Natural Sciences and Engineering
Research
Council of Canada.



\begin{references}

\bibitem{BrYo2} J. D. Brown and J. W. York, Jr., Phys. Rev. {\bf D 47},
1420 (1993).

\bibitem{BrYoRev} J. D. Brown and J. W. York, Jr.,  ``The path integral
formulation of gravitational thermodynamics",  preprint IFP-UNC-491,
TAR-UNC-043, CTMP/007/NCSU, gr-qc/9405024.

\bibitem{Te} C. Teitelboim, ``Action and entropy for extremal and
non-extremal black holes", preprint,  hepth/9410103, (1994).

\bibitem{CaTe} S. Carlip and C. Teitelboim, ``The off-shell black hole",
preprint,  IASSNS-HEP-93/84, UCD-93-34, gr-qc/9312002.

\bibitem{HaHoRo} S. W. Hawking, G. T. Horowitz, and S. F. Ross, ``Entropy,
area, and black hole pairs", preprint, DAMTP-R-94/26, UCSBTH-94-25,
gr-qc/9409013, (1994).

\bibitem{BrMaYo} J. D. Brown, E. A. Martinez, and J. W. York, Jr., Phys.
Rev. Lett., {\bf 66}, 2281 (1991); in {\it Nonlinear Problems in
Relativity and Cosmology}, edited by J. R. Buchler, S. L. Detweiler, and
J.R. Ipser (New York Academy of Sciences, New York, 1991).

\bibitem{FrNo} V. P. Frolov and I. Novikov, Phys. Rev. {\bf D 48}, 4545 (1993).

\bibitem{BaFrZe} A. O. Barvinsky, V. Frolov, and A. Zelnikov,
``Wavefunction of a black hole and the dynamical origin of entropy",
 preprint, Alberta-Thy-13-94, gr-qc/9404036, (1994).

\bibitem{LoWh} J. Louko and B. F. Whiting, ``Hamiltonian thermodynamics of the
Schwarzschild black hole", preprint, UF-RAP-94-13, WISC-MILW-94-TH-24,
gr-qc/9411017, (1994).

\bibitem{FrMa2} V. Frolov and E. A. Martinez, ``Action and Hamiltonian
for eternal black holes", preprint, Alberta-Thy-32-94, gr-qc/9411001,
(1994).

\bibitem{GiHa1} G. W. Gibbons and S. W. Hawking, Phys. Rev. {\bf D 15},
2752 (1977); S. W. Hawking, in {\it General Relativity}, edited by S. W.
Hawking and W. Israel (Cambridge University Press, Cambridge, 1979).

\bibitem{Yo} J. W. York, Jr., Phys. Rev. {\bf D 33}, 2092 (1986).

\bibitem{WhYo} B. F. Whiting and J. W. York, Jr., Phys. Rev. Lett., {\bf
61}, 1336 (1988).

\bibitem{GiHa2} G. W. Gibbons and S. W. Hawking, Commun. Math. Phys.
{\bf 66},  291 (1979).

\bibitem{IsSt} W. Israel and J. M. Stewart, in {\it General Relativity and
Gravitation. II}, edited by  A. Held (Plenum Press, New York, 1980).

\bibitem{FrTh} V. Frolov and K. S. Thorne, Phys. Rev. {\bf D 31}, 2125
(1989).

\bibitem{FrMa1} V. Frolov and E. A. Martinez, ``Eternal black holes and
quasilocal energy", Alberta-Thy-19-94, gr-qc/9405041,  in {\it
Proceedings of the Lake
Louise Winter Institute on Particle Physics and Cosmology}, edited by B.
Campbell and F. Khana, World Scientific, 1994.

\bibitem{ADM} R. Arnowitt, S. Deser, and C. W. Misner, in {\it
Gravitation: An Introduction to Current Research}, edited by L. Witten
(Wiley, New York, 1962).

\bibitem{BrYo1} J. D. Brown and J. W. York, Jr., Phys. Rev. {\bf D 47}, 1407
(1993).

\bibitem{ensembles} J. D. Brown, G. L. Comer, E. A. Martinez, J. Melmed,
B. F. Whiting, and J. W. York, Class. Quantum Grav. {\bf 7}, 1433 (1990).

\bibitem{MTW} C. W. Misner, K. S. Thorne and J. A. Wheeler, {\it
Gravitation}, W. H. Freeman, San Francisco, 1973.

\bibitem{Ha} G.Hayward, Phys. Rev. {\bf D 47}, 3275, (1993); G. Hayward and
K. Wong, Phys. Rev. {\bf D 46}, 620, (1992); Phys. Rev. {\bf D 47}, 4778,
(1993).

\bibitem{BrHa} D. Brill and G. Hayward,  Phys. Rev. {\bf D 50},  4914
(1994).

\bibitem{Ku76} K. V. Kucha\v{r}, J. Math. Phys. {\bf 17}, 777 (1976);
{\bf 17}, 792 (1976); {\bf 17}, 801 (1976).

\bibitem{Ba} J. M. Bardeen, in {\it Black Holes}, edited by C. DeWitt and B. S.
DeWitt, Gordon and Breach Science Publishers, New York, 1973.

\bibitem{UmTa} H. Umezawa and Y. Takahashi, Collective Phenomena {\bf 2},
55 (1975); H. Umezawa, {\it Advanced Field Theory}, American Institute of
Physics, New York, 1993.

\bibitem{Is} W. Israel, Phys. Lett. {\bf 57A}, 107 (1976).

\bibitem{tfd} R. Laflamme, Nucl. Phys. {\bf B324}, 233 (1989);
N. Sanchez and B. F. Whiting, Nucl. Phys. {\bf B283}, 605 (1987);
B. F. Whiting, in {\it Proceedings of the Workshop on Thermal Field
Theories and their applications}, edited by K. L. Kowalski, N. P.
Landsman, and Ch. G. van Weert, Physica A, North Holland (1988).

\bibitem{HaLo} G. Hayward and J. Louko, Phys. Rev. {\bf D 42}, 4032, (1990).

\bibitem{Ku1} K. V. Kucha\v{r}, Phys. Rev. {\bf D 50}, 3961 (1994).

\bibitem{HaHa} J. B. Hartle and S. W. Hawking, Phys. Rev. {\bf D 13},
2188 (1976).

\bibitem{Ja} T. Jacobson, ``A note on Hartle-Hawking vacua",  preprint,
gr-qc/9407022, (1994).




\end{references}
\end{document}